\documentclass[twocolumn,prb,floatfix,amssymb,aps]{revtex4}
\usepackage{graphicx}

\begin{document}

\hsize\textwidth\columnwidth\hsize\csname@twocolumnfalse\endcsname

\title{Insulating charge density wave for a half-filled SU(N)
Hubbard model with an attractive on-site interaction in one
dimension}
\author{Jize Zhao and Kazuo Ueda}
\affiliation{Institute for Solid State Physics, University of
Tokyo, Kashiwa, Chiba 277-8581, Japan}
\author{Xiaoqun Wang}
\affiliation{Department of Physics, Renmin University of China,
Beijing 100872, China}
\date{\today}
\begin{abstract}
We study a one-dimensional SU(N) Hubbard model with an attractive
on-site interaction and $N>2$ at half-filling on the bipartite
lattice using density-matrix renormalization-group method and a
perturbation theory. We find that the ground state of the SU(N)
Hubbard model is a charge density wave state with two-fold
degeneracy. All the excitations are found to be gapful, resulting
in an insulating ground state, on contrary to that in the SU(2)
case. Moreover, the charge gap is equal to the Cooperon gap, which
behaves as $-2Nt^2/(N-1)U$ in the strong coupling regime. However,
the spin gap $\Delta_{s}$ and the quasiparticle gap $\Delta_{1}$
as well open exponentially in the weak coupling region, while in
the strong coupling region, they linearly depend on $U$ such that
$\Delta_{s}\sim -U(N-1)$ and $\Delta_{1}\sim -U(N-1)/2$.
\end{abstract}

\pacs{71.10.Fd, 71.10.Pm, 71.30.+h}
\maketitle

\section{INTRODUCTION}
Correlation effects in electronic systems have been of long-term
interest in the condensed matter physics. In recent years,
important progress has been made experimentally in ultra-cold atomic
systems where strong correlation leads to some novel physical
phenomena. In particular, interacting fermionic atoms can be trapped
in an optical lattice\cite{kohl05,matrin}. More interestingly, the interaction
in the ultra-cold atomic systems is tunable through the Feshbach resonance, which
allows for a full exploration of various fundamental properties of
strongly correlated models. Moreover, the nuclear
spin of the atoms can be larger than the electronic spin.
One could expect richer physics 
induced by the degree of freedom of higher spins\cite{HO1, WU1, WU2, TU1}.
Therefore, one would like naturally to generalize the Hubbard model
with two spin components to the one with $N$-components.

In this paper, we investigate low energy properties of a one-dimensional
half-filled SU(N) Hubbard model\cite{AFFLECK1} with an attractive on-site
interaction and $N>2$. This is a generalization of our previous
work in which we have focused on the SU(4) case \cite{ZHAO1}.
The Hamiltonian of the one-dimensional SU(N) Hubbard model is
represented by
\begin{equation}
\mathcal{H}=-t\sum_{i=1}^{L}\sum_{\sigma}(\hat{c}^{\dagger}_{i\sigma}\hat{c}_{i+1\sigma}+h.c.)+\\
\frac{U}{2}\sum_{i=1}^{L}\sum_{\sigma\ne\sigma^{'}}\hat{n}_{i\sigma}\hat{n}_{i\sigma^{'}},
\label{HAM}
\end{equation}
where $t>0$ is the hopping matrix set as the energy unit, $U<0$
the coupling constant of the attractive on-site interaction, $L$
the number of lattice sites, $\sigma$ and $\sigma^{'}$ the spin
indices, which take the values $(N-1)/2$, $(N-1)/2-1$, $\cdots$,
$1-(N-1)/2$, $-(N-1)/2$, respectively.
$\hat{c}^{\dagger}_{i\sigma}$ and $\hat{c}_{i\sigma}$ denote the
creation and annihilation operators, respectively, for a particle
with spin $\sigma$ at the site $i$ and
$\hat{n}_{i\sigma}=\hat{c}^{\dagger}_{i\sigma} \hat{c}_{i\sigma}$
indicates the operator of the particle number. The Hamiltonian (1)
has U(1)$\otimes$SU(N) symmetry\cite{ASSARAF1}. The generator for
the U(1) symmetry is $\mathcal{N}=\sum_{i\sigma}\hat{n}_{i\sigma}$
and the U(1) symmetry implies that the particle number
$\mathcal{N}$ is conserved. The generator for the SU(N) symmetry
is $S^{A}=\sum_{i\sigma\sigma^{'}}
\hat{c}^{\dagger}_{i\sigma}\mathcal{T}^{A}_{\sigma\sigma^{'}}\hat{c}_{i\sigma^{'}}$,
where $\mathcal{T}^{A}$ are the generators of the SU(N) group in
its fundamental representation. These symmetries are useful for
simplifying numerical calculations and classifying excitations for
the present system.

Although this model is exactly solvable for $N=2$\cite{LIEB1}, it
seems that physical features obtained for $N=2$ are not immediately
applicable to more general cases with $N>2$\cite{CHOY1}. Recently, this model
with $N>2$ and $U>0$ has been studied by using several analytic as
well as numerical approaches. At half-filling, a
renormalization-group analysis \cite{AFFLECK1} shows that both the
charge and spin coupling constants are renormalized to a large
value, resulting in gaps in both sectors. An analytic perturbative
renormalization group treatment in the fermionic representation
alternatively gives rise to a gapful spectrum for all $N>2$ and
$U>0$\cite{SZIRMAI1}. Moreover, by employing bosonization method
and quantum Monte Carlo simulations, Assaraf et al found for a
$1/N$ filling that the spin excitation is gapless for any positive
$U$, whereas the charge excitation is gapful only for $U>U_c$
where $U_c\neq 0$. However, Buchta et al obtained gapless spin as
well as gapful charge excitations from accurate density-matrix
renormalization group(DMRG) calculations with $N=3, 4$ and $5$ for
any $U>0$.

For the attractive interaction, on the other hand, it is known
that the one-dimensional attractive half-filled SU(2) Hubbard
model is described by a Luther-Emery liquid model, in which the
charge excitation is gapless, whereas the spin excitation is
gapful. By the hidden SU(2) transformation, the SU(2) Hubbard
model with $U$ can be mapped to the one with $-U$. However, for
the SU(N) Hubbard model with $N>2$ such a mapping does not exist
so that one cannot obtain any insight into the low-energy
properties for the attractive case through the mapping from the
repulsive case. In this paper, we will show that the SU(N) Hubbard
model at half filling with an attractive interaction belongs to a
different universality class from the SU(2) one and all the
excitations are gapful. We expect that our findings
are not only of fundamental interest, but also useful for
experimentalists, since the attractive SU(N) Hubbard model may be
possibly realized by experiments in ultra-cold atomic systems.

The rest of this paper is organized as follows.
In Sec. II, we analyze low-energy properties of the Hamiltonian (\ref{HAM})
in the strong coupling limit $(N-1)|U|\gg t$ first and then in the weak coupling
limit $(N-1)|U|\ll t$ by the perturbation treatments. Particularly we discuss
the dependence of charge gap, Cooperon gap, spin gap and quasiparticle gap on $U$ in both regimes.
In Sec. III, we present our numerical results which are obtained from the DMRG calculations
and compare them with analytic behavior in both weak and strong coupling regime.
A summary is finally given in Sec. IV.

\section{PERTURBATION CALCULATIONS}
\begin{figure}[ht]
\includegraphics[width=7.8cm,angle=0]{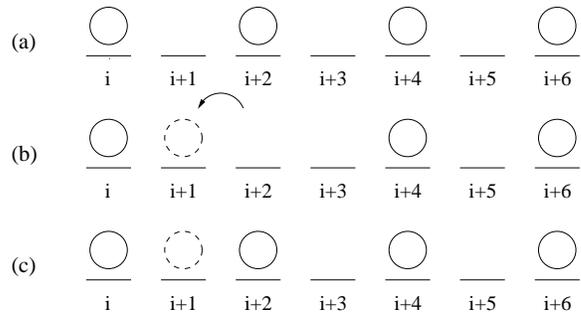}
\caption{Schematic illustrations of the configurations, and the
empty circle here represents one SU(N) singlet formed by $N$
particles with different $\sigma$. (a) For the CDW ground state.
Shifting all the circles by one site, one can obtain the other one
of two-fold degenerate ground states. (b) For the charge
excitation: the SU(N) singlet at site $i+2$ shifts to the site
$i+1$. (c) For the Cooperon excitation: one SU(N) singlet is added
to the site i+1.} \label{fig0}
\end{figure}

In this section, we study the low-energy properties of the
Hamiltonian (\ref{HAM}) by a perturbation theory. 
For this purpose, we rewrite it as
\begin{eqnarray}
\mathcal{H}=\mathcal{H}_t+\mathcal{H}_u,
\end{eqnarray}
where
$\mathcal{H}_t=-t\sum_{i\sigma}(\hat{c}^{\dagger}_{i\sigma}\hat{c}_{i+1\sigma}+h.c.)$
is the hopping term and
$\mathcal{H}_{u}=(U/2)\sum_{i\sigma\ne\sigma^{'}}\hat{n}_{i\sigma}\hat{n}_{i\sigma^{'}}$
the on-site interaction. We start with the on-site interaction
part $\mathcal{H}_{u}$. Since the on-site interaction is
attractive, $N$ particles with different $\sigma$ tend to stay on
one site and form a SU(N) singlet. The energy of the SU(N) singlet
is $UN(N-1)/2$. In the half-filling case where $N/2$ particles per
site, the ground states of $\mathcal{H}_u$ are highly degenerate,
involving half of the lattice sites occupied by the SU(N) singlets
and the other half being empty. In the strong coupling region,
i.e. $|U|(N-1)\gg t$, $\mathcal{H}_{u}$ is taken as the zeroth
order Hamiltonian, while $\mathcal{H}_t$, being the order of $Nt$,
is regarded as a perturbation. Up to the second order, we obtain
the first order effective Hamiltonian
\begin{eqnarray*}
\mathcal{H}_{eff}^{(1)}=P\mathcal{H}_tP=0 ,
\end{eqnarray*}
where $P$ is a projection operator which restricts the effective
Hamiltonian in the subspace spanned by the ground states of
$\mathcal{H}_{u}$. The second order effective Hamiltonian
reads
\begin{eqnarray}
\mathcal{H}_{eff}^{(2)} & = & P\mathcal{H}_{t}\frac{1}{E_0-\mathcal{H}_u}(1-P)\mathcal{H}_tP \nonumber\\
                        & = & \frac{2t^2}{(N-1)U}P\sum_{i}\hat{n}_{i}P\nonumber\\
                        &   & - \frac{2t^2}{N(N-1)U}P\sum_{i}\hat{n}_{i}\hat{n}_{i+1}P,
\label{HEFF}
\end{eqnarray}
where $\hat{n}_{i}=\sum_{\sigma}\hat{n}_{i\sigma}$ the number
operator at the site $i$. The degeneracy of the ground state of
$\mathcal{H}_u$ is thus lifted by $\mathcal{H}_t$. At half
filling, one can easily obtain the ground state energy correction per site
which is given as $Nt^2/(N-1)U$ by the first term of Eq.
(\ref{HEFF}). Since the second term introduces an effective
repulsion interaction between the particles at nearest
neighbor(NN) sites, it results in a charge density wave (CDW)
state in which the SU(N) singlet and the empty site occur
alternatively to exhibit a long range order resulting from 
translational symmetry breaking. It turns out that the ground
state is two-fold degenerate. The configuration of the
ground state is schematically shown in Fig. \ref{fig0}(a).

It is well-known that both the charge and Cooperon excitations for
the attractive SU(2) Hubbard model are gapless. However for the
SU(4) case, it has been shown\cite{ZHAO1} that the charge and
Cooperon excitations are gapful and equal to each other at
half-filling. In the following, one can see that this conclusion
is also valid for other attractive SU(N) Hubbard models with $N>2$.
In particular, the charge gap and Cooperon gap can be
easily derived from the effective Hamiltonian (\ref{HEFF}). The
charge gap $\Delta_c$ is defined as the lowest excitation in the
SU(N) singlet subspace as follows
\begin{eqnarray}
\Delta_c=E_1(L,NL/2,0)-E_0(L,NL/2,0), \label{DELTAC}
\end{eqnarray}
where $E_0(L,M,S)$ is the ground state energy in the spin-$S$
channel with $L$ sites and $M$ particles, and $E_n(L,M,S)$ the
$n$-th excitation energy. The gap for Cooperon excitations for the
SU(N) Hubbard model is given from the energy
difference between states by adding $N$ particles or $N$ holes to
the system, which is defined as
\begin{eqnarray}
\Delta_N & = & \frac{1}{2}\left[E_0(L,\frac{NL}{2}+N,0)
           +E_0(L,\frac{NL}{2}-N,0)\right]\nonumber\\
         &   & -E_0(L,\frac{NL}{2},0).
\end{eqnarray}
The charge gap $\Delta_c$ is given by shifting one of SU(N)
singlets in the ground state configuration to its nearest neighbor
site as shown in Fig. \ref{fig0}(b). Then one has
\begin{eqnarray}
\Delta_c=-\frac{2Nt^2}{(N-1)U}. \label{DCHARGE}
\end{eqnarray}
Similarly, 
one obtains the SU(N) Cooperon gap
\begin{eqnarray}
\Delta_N=-\frac{2Nt^2}{(N-1)U}.
\end{eqnarray}
which is shown in Fig. \ref{fig0}(c).
One can see that $\Delta_c=\Delta_N$ for all $N>2$, which was
previous shown for the $N=4$ case\cite{ZHAO1}. In the large-$N$
limit,
\begin{eqnarray}
\Delta_c=\Delta_N=-\lim_{N\rightarrow\infty}\frac{2Nt^2}{(N-1)U}=-\frac{2t^2}{U}.
\end{eqnarray}
We note that the difference for the low-lying excitations between
$N=2$ and $N>2$ results from the effective interaction
between the singlets at the NN sites as involved in effective
Hamiltonian (\ref{HEFF}). For the $N=2$ case, one has the
effective Hamiltonian\cite{EMERY1}
\begin{eqnarray}
\mathcal{H}^{(2)}_{eff,su(2)} & = &
\frac{2t^2}{U}P\sum_{i\sigma}\hat{n}_{i\sigma}P \label{HEFFSU2}\\
     &  & - \frac{t^2}{U}P\sum_{\langle{ij}\rangle\sigma}
             (\hat{n}_{i\sigma}\hat{n}_{j\sigma}-\hat{c}^{+}_{i\sigma}\hat{c}^{+}_{i\bar{\sigma}}
             \hat{c}_{j\bar{\sigma}}\hat{c}_{j\sigma})P. \nonumber
\end{eqnarray}
Compared to the effective Hamiltonian (\ref{HEFF}), one has
additionally a pair hopping term, which involves the same
amplitude as the NN repulsion term and eventually destroys the CDW
long range order for $N=2$. On the other hand, for
$N>2$ cases, although a similar hopping term emerges at the $N$-th
order perturbation, it has a smaller amplitude than the NN
repulsion term so that one can has a stable CDW ground state.

Now we turn to study the spin and quasiparticle excitations. 
The spin gap is defined in correspondence to the
lowest excitation with different spin quantum number from the
ground state. The quasiparticle gap is defined as a
energy change by adding one particle or hole to the system. Since
the ground state is CDW, in order to obtain these two gaps, we
resort to the following Hartree-Fock(HF) approximation.

\begin{eqnarray}
\hat{n}_{i\sigma}\hat{n}_{i\sigma^{'}}\simeq n_{i\sigma}\langle{\hat{n}_{i\sigma^{'}}}\rangle+
\langle{\hat{n}_{i\sigma}}\rangle\hat{n}_{i\sigma^{'}}-\langle{\hat{n}_{i\sigma}}\rangle
\langle{\hat{n}_{i\sigma^{'}}}\rangle,
\end{eqnarray}
where $\langle{\hat{n}_{i\sigma}}\rangle=n_0 +(-1)^{i}\delta{n}$,
and $\delta{n}$ is the order parameter, $\langle\cdots\rangle$ is
the average over the ground state. The HF Hamiltonian then reads
\begin{eqnarray}
\mathcal{H}^{HF}&=&-t\sum_{i\sigma}(\hat{c}_{i\sigma}^{\dagger}\hat{c}_{i+1\sigma}+h.c.)
+\frac{U}{2}\sum_{i\sigma\ne\sigma^{'}}(\hat{n}_{i\sigma}\langle{\hat{n}_{i\sigma^{'}}}\rangle\nonumber\\
&+&
\langle{\hat{n}_{i\sigma}}\rangle\hat{n}_{i\sigma^{'}}-\langle{\hat{n}_{i\sigma}}\rangle
\langle{\hat{n}_{i\sigma^{'}}}\rangle). \label{HHF}
\end{eqnarray}
At half-filling, one has $n_0=\frac{1}{2}$ and
$\sum_{l\sigma}\hat{n}_{l\sigma}=\frac{LN}{2}$. Eq. (\ref{HHF})
can be further simplified as
\begin{eqnarray}
\mathcal{H}^{HF}&=&-t\sum_{i\sigma}(\hat{c}^{\dagger}_{i\sigma}\hat{c}_{i+1\sigma}+h.c.)\nonumber\\
&+&U(N-1)\delta{n}\sum_{i\sigma}(-1)^{i}\hat{n}_{i\sigma}+{\rm
const}.
\end{eqnarray}
Introducing $\hat{a}_{l\sigma}=\hat{c}_{2l\sigma}$,
$\hat{b}_{l\sigma}=\hat{c}_{2l+1\sigma}$ and taking a Fourier
transformation
$\hat{a}_{l\sigma}=\sqrt{\frac{2}{L}}\sum_{k}\hat{a}_{k\sigma}e^{ikl}$,
$\hat{b}_{l\sigma}=\sqrt{\frac{2}{L}}\sum_{k}\hat{b}_{k\sigma}e^{ikl}$,
we obtain
\begin{eqnarray}
\mathcal{H}^{HF}&=&-t\sum_{k\sigma}((1+e^{-ik})\hat{a}^{\dagger}_{k\sigma}\hat{b}_{k\sigma}+h.c.)\\
&+&U(N-1)\delta
n\sum_{k\sigma}(\hat{a}^{\dagger}_{k\sigma}\hat{a}_{k\sigma}-\hat{b}^{\dagger}_{k\sigma}\hat{b}_{k\sigma})+{\rm
const}.\nonumber
\end{eqnarray}
Diagonalizing this HF Hamiltonian, we can obtain two bands for
each spin species with the following dispersion
\begin{eqnarray}
\omega_{k\sigma}=\pm \sqrt{\Delta_{1}^{2}+4t^2\cos^{2}{\frac{k}{2}}},
\end{eqnarray}
where $\Delta_{1}=-(N-1)U\delta n$ is the quasiparticle gap. Under
the HF approximation, one has that
$\Delta_{c}=\Delta_{s}=2\Delta_{1}$. By solving the
self-consistent equation for $\delta n=\langle
\hat{a}^{\dagger}_{l\sigma}\hat{a}_{l\sigma}-\hat{b}^{\dagger}_{l\sigma}\hat{b}_{l\sigma}\rangle$
one can determine the order parameter $\delta n$. In the weak
coupling limit $t\ll |U|(N-1)$, we can solve it approximately and
obtain
\begin{eqnarray}
\Delta_{1}\simeq 2\pi te^{\frac{2\pi t}{U(N-1)}}.
\end{eqnarray}
In the strong coupling limit, one can put $\delta{n}\simeq 0.5$, and then
obtain the quasiparticle gap
\begin{eqnarray}
\Delta_{1}\simeq -U(N-1)/2 ,
\label{D1}
\end{eqnarray}
and the spin gap
\begin{eqnarray}
\Delta_{s}\simeq -U(N-1) .
\label{DS}
\end{eqnarray}

In the HF approximation, since $\Delta_c=2\Delta_1$, one obtains
$\Delta_c\simeq -U(N-1)$ which is inconsistent with the results
Eq. (\ref{DCHARGE}) obtained from the strong coupling perturbation
theory. This is because in the presence of the strong attractive
on-site interaction between particles, the single particle picture is not
valid for the charge excitations, in which $N$ particles forming a
SU(N) singlet shift as a whole to the NN site, as illustrated in
Fig. \ref{fig0}(b). However, we would expect that the HF results are
still qualitatively correct for weak coupling limit.

\section{NUMERICAL CALCULATIONS}
In this section, we present our numerical results for $N=3, 4, 5$
and $6$ cases and compare them with those analytic predictions
from the perturbation theories and the HF approximation. Our
numerical results are obtained from a large scale DMRG
computation.\cite{WHITE1,PESCHEL1,SCHOLLWOCK1} It is well known
that DMRG method is the most powerful numerical tool for accurate
exploration on low-energy properties of one dimensional systems at
zero temperature. However, it is nontrivial to apply this method
to the SU(N) Hubbard model. On one hand, there is a large number
of degrees of freedom per site. For instance, the degree of
freedom per site is 32 when $N=5$. On the other hand, although a
open boundary condition (OBC) in DMRG calculations may lead to
more accurate results than a periodic boundary condition (PBC),
while the emergence of edge states under the OBC makes the
calculation of various excitations practically more difficult.

For these reasons, we have made some additional considerations to
the standard DMRG algorithm as follows. When $N=3$, i.e. SU(3)
case, one has 8 degrees of freedom per site. In this case, we use
the PBC with keeping about $2000\sim 3500$ optimal states in the
DMRG procedure and two sites are added in order to enlarge the
chain length at each DMRG step. The maximal truncation error is of
the order $10^{-5}$. Similarly, the charge gap for $N=4$ is
calculated with at most 1600 states kept. The other excitations
are calculated with the OBC. At each step we add one lattice site
to the chain, breaking the lattice site into two pseudo sites. To
obtain accurate results, we preselect some specific chain lengths,
and perform sweeping at these preselected lengths, the final
results are got by extrapolating the results at the preselected
lengths to the thermodynamic limit. To avoid cumbersome edge
excitations when the OBC is employed, the preselected lengths are
odd instead of even. Correspondingly the definitions of the
excitations are changed. For example, the Cooperon gap for $N=4$
is redefined as $\Delta_{N}=E_0(L,2L+6,0)-E_0(L,2L+2,0)-6U$, where
the particle-hole symmetry is taken into account and odd $L$ are
used.

\subsection{Energy correction and degeneracy of the ground states}
\begin{figure}[ht]
\includegraphics[width=7.0cm,angle=0]{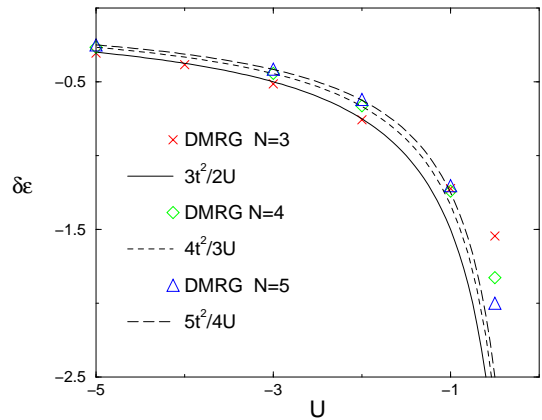}
\caption{Ground state energy correction per site obtained from
both the DMRG calculation and the strong coupling perturbation
theory $\mathcal{H}^{(2)}_{eff}$ for N=3, N=4 and N=5 are shown as
a function of $U$. } \label{fig1}
\end{figure}

The analysis based on the strong coupling perturbation theory in
the above section shows that the hopping term, lifting the
degeneracy of $\mathcal{H}_u$, results in an energy correction per
site $Nt^{2}/(N-1)U$. To verify this prediction, we calculate the
energy corrections for $N=3, 4$ and $5$ using DMRG method. Both
numerical and analytic results are shown in Fig. \ref{fig1}.
One can see that the energy correction is negative and decreases
monotonically with respect to increasing of $U$. In the strong
coupling region, the results given by the perturbation theory
agree well with the DMRG data. For $N=3$, the deviation between
analytic and numerical results is within the numerical accuracy
for $U<-2$. With increasing of $N$, the deviation becomes smaller
and smaller. This implies that the perturbation theory gives rise
to better results for larger $N$ for the strong coupling regime.
In the weak coupling region, the analytic results from the
effective Hamiltonian (\ref{HEFF}) severely deviates from the DMRG
results. This is reasonable since in the weak coupling region,
$\mathcal{H}_t$ is no longer a small perturbation. Nevertheless,
when $N$ increases, the valid region of the strong coupling
perturbation theory increases simultaneously.

\begin{figure}[ht]
\includegraphics[width=7.0cm,angle=0]{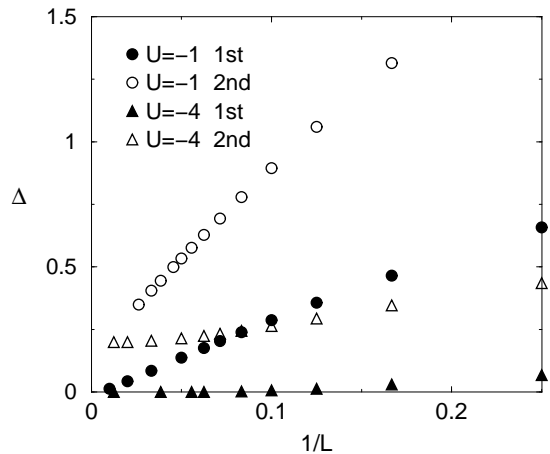}
\caption{Two lowest excitation gaps in the SU(N) singlet subspace
for $N=3$ case with two different values of $U$. In the thermodynamic
limit, the first excitation energy becomes degenerate with the ground
state, while the energy difference between the second excitation
and the ground state remains finite, indicating a gapful excitation.}
\label{fig2}
\end{figure}

\begin{figure}[ht]
\includegraphics[width=7.0cm,angle=0]{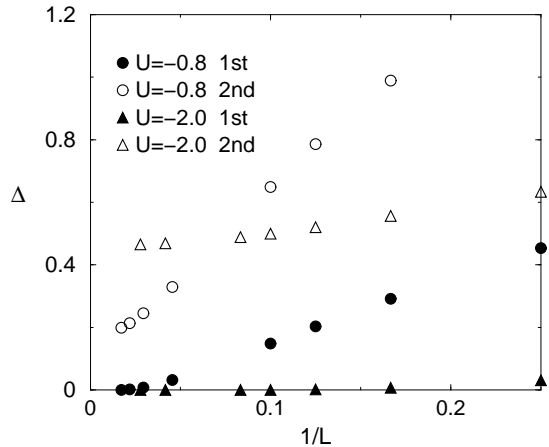}
\caption{Two lowest excitation gaps in the SU(N) singlet subspace
for $N=4$ case with two different values of $U$. In the
thermodynamic limit, the first excitation energy becomes degenerate with
the ground state, while the energy difference between the second
excitation and the ground states remains finite, indicating a gapful 
excitation.} \label{fig3}
\end{figure}

We have examined the symmetry properties of the ground state. In
the thermodynamic limit, the ground state is two-fold degeneracy. For
a finite $L$, the ground state is a SU(N) singlet, and there is
another SU(N) singlet state just above the ground state. When the
chain length $L$ is increased, the energy difference between these
two states decreases and eventually vanishes in the large-$L$
limit. This is consistent with the prediction from the effective 
Hamiltonian (\ref{HEFF}). Moreover, for a given chain length $L$, we find that the
energy difference between two states for the Hamiltonian
(\ref{HAM}) decrease rapidly with increasing $|U|$, since the
effective Hamiltonian Eq. (\ref{HEFF}) becomes more accurate for
larger $|U|$. In Fig. \ref{fig2} and \ref{fig3}, we demonstrate
this feature with two different values of $U$ under OBC for $N=3$
and $4$, respectively. Note that the energy difference for the
second excitation, which is finite in the large-$L$ limit,
corresponds to edge excitations rather than bulk excitations.

\subsection{Charge gaps and Cooperon gaps}
\begin{figure}[ht]
\includegraphics[width=7.0cm,angle=0]{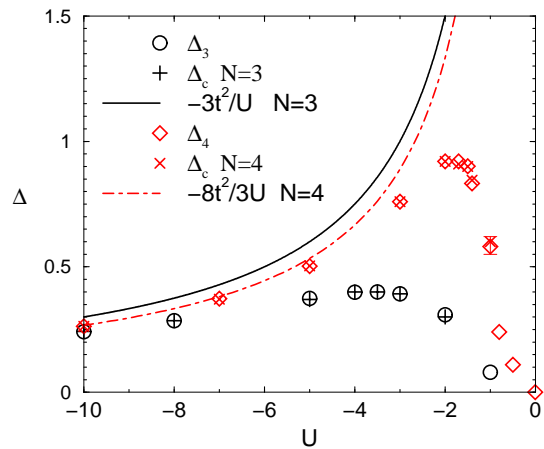}
\caption{The Cooperon and charge gaps are shown for $N=3$ and $4$.
The asymptotic behavior from the effective Hamiltonian(\ref{HEFF})
(curves) are also shown for comparison.} \label{fig4}
\end{figure}

The prediction that the charge gap $\Delta_c$ is equal to the
Cooperon gap $\Delta_N$ is confirmed numerically for N=4\cite{ZHAO1}. 
However, the strong coupling perturbation theory gives rise to
$\Delta_c=\Delta_N$ for general $N$, it is still interesting to
examine numerically whether this equation is valid for both even and odd N and all
regimes of the coupling constant $U$\cite{BUCHTA1}. Fig.
\ref{fig4} demonstrates the charge and Cooperon gaps for $N=3$ as
comparison to $N=4$ as a function of $U$.
One can see that both charge and Cooperon gaps for both $N=3$ and
$N=4$ are finite for all $U<0$. When $|U|$ is increased, these
gaps increase first in the weak coupling region. After they reach
their maxima, they decrease following the asymptotic behavior
$-\frac{2Nt^2}{(N-1)U}$ resulting from the strong coupling perturbation theory. We
would mention that even beyond the valid range of the perturbation
theory, our numerical data still show that the charge gaps are
equal to the corresponding Cooperon gaps for both odd $N (=3)$ and
even $N (=4)$. Therefore, one may conclude that
$\Delta_c=\Delta_N$ for all $N$ and $U<0$.

\begin{figure}[ht]
\includegraphics[width=7.0cm,angle=0]{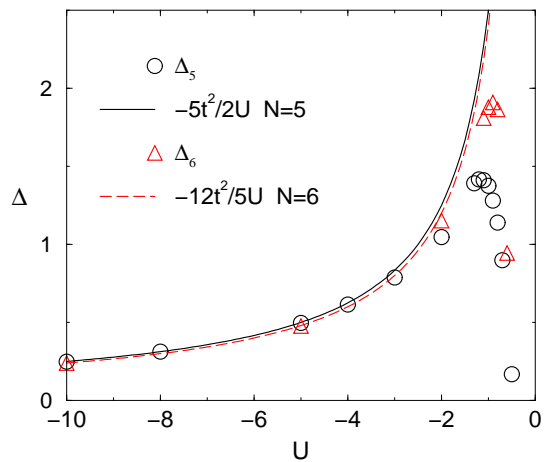}
\caption{The Cooperon gaps and the corresponding asymptotic behaviors as a function of $U$ for N=5 and 6
are shown.}
\label{fig6}
\end{figure}

Moreover we calculate the Cooperon gaps for $N=5$ and $6$ in order
to explore the large-$N$ behavior of the charge as well as
Cooperon gaps. The results of the Cooperon gap for $N=5$ and $6$
are shown in Fig. \ref{fig6}.
Comparing Fig. \ref{fig4} and \ref{fig6}, one can see that the
larger $N$ is, the closer  Cooperon gaps to that derived from the
effective Hamiltonian (\ref{HEFF}). For $N=3$ as shown in Fig.
\ref{fig4}, the deviation of the asymptotic behavior from the DMRG
data is visible even at $U=-10$, while for $N=4$ case, the
deviation is within the numerical accuracy up to $U\simeq -5$.
For $N=5$ and $6$, the good agreement can be seen up
to $U\simeq -3$ and $U\simeq -2$, respectively. Furthermore, as
$N$ increases, the height of the peak of the Cooperon gaps
increases and the position of the peak shifts toward $U=0$. These
features can be easily understood since the criteria for the
strong coupling perturbation $t\ll |U|(N-1)$ depends on $|U|(N-1)$
rather than $U$ only. The DMRG results here also verify that the
charge and Cooperon gap behaves asymptotically as
$-\frac{2t^2}{U}$ in the large-$N$ limit.

\subsection{Quasiparticle gap and spin gap}
In the following, we discuss the quasiparticle and spin gaps. In
order to improve numerical efficiency, we need to carry out DMRG
calculations in a subspace which can be obtained by decomposing
the irreducible representations of SU(N) into the irreducible
representation of SO(3)\cite{HAMERMESH1,YAMASHITA1}. Numerically,
one can determine the irreducible representations for short chains
by varying $z$-component. Table \ref{TABLE1} shows the
decomposition of some irreducible representations of SU(3) to
SO(3).
\begin{table} \caption{Decomposition of
some irreducible representations of SU(3) to SO(3).}
\begin{ruledtabular}
\begin{tabular}{cccc}
$SU(3)$&$SO(3)$&$\nu$\\
\hline
$[1^3]$&$0$&1\\
$[2^{1}1^{1}]$&$2\oplus{1}$&8\\
$[3^1]$&$3\oplus{1}$&10\\
\hline
$[1^1]$&$1$&3\\
$[2^2]$&$2\oplus{0}$&6 \label{TABLE1}
\end{tabular}
\end{ruledtabular}
\end{table}

\begin{table}[ht]
\caption{Decomposition of some irreducible representations of SU(4) to SO(3).}
\begin{ruledtabular}
\begin{tabular}{cccc}
$SU(4)$&$SO(3)$&$\nu$\\
\hline
$[1^4]$&$0$&1\\
$[2^2]$&$4\oplus{2}\oplus{2}\oplus{0}$&20\\
$[2^{1}1^{2}]$&$3\oplus{2}\oplus{1}$&15\\
$[4^1]$&$0\oplus{2}\oplus{3}\oplus{4}\oplus{6}$&35\\
$[3^{1}1^{1}]$&$1\oplus{1}\oplus{2}\oplus{3}\oplus{3}\oplus{4}\oplus{5}$&45\\
\hline
$[1^1]$&$\frac{3}{2}$&4\\
$[2^{2}1^{1}]$&$\frac{7}{2}\oplus\frac{5}{2}\oplus\frac{3}{2}\oplus\frac{1}{2}$&20\\
$[3^{1}1^{2}]$&$\frac{9}{2}\oplus\frac{7}{2}\oplus\frac{5}{2}\oplus\frac{5}{2}\oplus\frac{3}{2}\oplus\frac{1}{2}$&36
 \label{TABLE2}
\end{tabular}
\end{ruledtabular}
\end{table}

\begin{figure}[ht]
\includegraphics[width=7.0cm,angle=0]{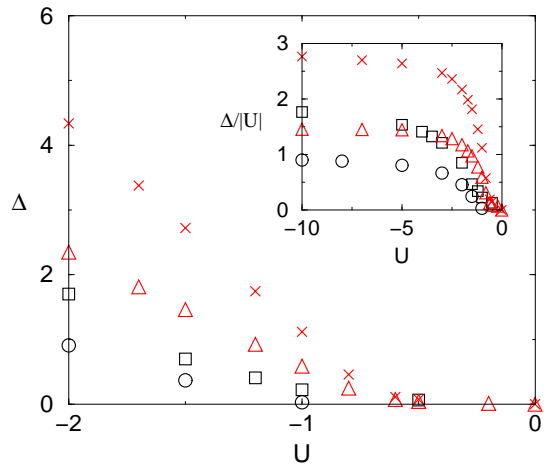}
\caption{The quasiparticle gap and spin gap for N=3, 4 are shown in the figure.
For N=3, the quasiparticle gap($\small{\bigcirc}$) and spin gap($\square$) are shown by the black symbols.
For N=4, the quasiparticle gap($\triangle$) and spin gap($\times$) are shown by red symbols.}
\label{fig7}
\end{figure}

For $N=3$ that the spin excitation belongs to the
representation $[2^{1}1^{1}]$ and it is 8-fold degenerate,
while the quasiparticle excitation belongs to the representation
$[1^{1}]$ and it is 3-fold degenerate. Table \ref{TABLE2} shows
the decomposition of some irreducible representations of SU(4) to
SO(3). In this case, the spin excitation belongs to the
representation $[2^{1}1^{2}]$ and its degeneracy is 15-fold,
whereas the quasiparticle excitation belongs to the representation
$[1^1]$ and its degeneracy 4-fold. Fig. \ref{fig7} shows the quasiparticle
and spin gaps for $N=3$ and $4$ as well.
In weak coupling region, one can see that both the quasiparticle
and spin gaps open exponentially, as predicted by the HF approximation,
although we cannot determine precise critical behavior for the
opening of the gaps from our numerical data due to limited
numerical accuracy. In the strong coupling region, both the
quasiparticle gap and spin gap depend linearly on $U$ as seen in
the inset. This is also consistent with the HF results. Moreover, one
also see that the ratio $\Delta_{1}/|U|$ approaches 1 for $N=3$
asymptotically, while it approaches 1.5 for $N=4$. These asymptotic
behaviors qualitatively coincide with Eq. (\ref{D1}) and
(\ref{DS}) and the relation $\Delta_s=2\Delta_1$ derived from the
HF approximation. For finite $U$, however, the relation is just approximately
valid since there remains residual interaction between
quasiparticles.

\section{CONCLUSIONS}
In this paper, we have investigated low-energy properties of the
SU(N)(N$>$2) Hubbard model with attractive on-site interaction at
half-filling. By using the perturbation theory and DMRG method, we
show that the ground state is a CDW state with two-fold
degeneracy. The CDW long range order reflects the broken translational 
symmetry. On one hand, the strong coupling perturbation theory
predicts that both the charge excitations and Cooperon excitations
are gapful and equal to each other with the asymptotic behavior
$-2Nt^{2}/(N-1)U$. Combining this with numerical results for
$N=3, 4, 5$ and $6$, we conclude that the charge gap is equal to the
Cooperon gap for all $U<0$ and $N>2$. In the large N limit, both
the charge and Cooperon gaps behaves as $-2t^{2}/U$. 
Considering the CDW ground state, we
obtain qualitatively correct behavior for the spin and
quasiparticle gaps from HF approximation, which are confirmed by
our numerical data. In the weak coupling region, they open
exponentially, whereas in the strong coupling region, they depend
linearly on $U$ with the spin gap $\Delta_{s}\sim -(N-1)U$ and
quasiparticle gap $\Delta_{1}\sim -(N-1)U/2$.

Our findings indicate that at half filling, the SU(N)(N$>$2)
Hubbard model belongs to a different universality class from the
SU(2) case. This can be easily understood by observing the
difference between the effective Hamiltonians for the SU(N)(N$>$2)
and the SU(2) cases. In particular, the effective Hamiltonian
which for the SU(2) case involves an effective hopping term has
the same order as the effective repulsion interaction for
particles between nearest-neighbor sites. The hopping term
destroys the CDW order. For $N>2$, the repulsive
interaction dominates over the effective hopping term.

J. Zhao would thank H. Tsunetsugu for helpful discussion. X.Q.
Wang is supported in part by NFC2005CB32170X, and NSFC10425417
$\&$ C10674142.

\vfill
\end{document}